\documentstyle{article}

\catcode`\@=11

   \renewcommand{\section}%
   {\setcounter{equation}{0}\@startsection {section}{1}{\z@}{-3.5ex plus -1ex
   minus -.2ex}{2.3ex plus .2ex}{\Large\bf}}

   \newcommand{\beq}{\begin{equation}}
   \newcommand{\eeq}{\end{equation}}
   \newcommand{\beqs}{\arraycolsep1.5pt\begin{eqnarray}}
   \newcommand{\eeqs}{\arraycolsep5pt\end{eqnarray}}
   \newcommand{\beqsn}{\arraycolsep1.5pt\begin{eqnarray*}}
   \newcommand{\eeqsn}{\end{eqnarray*}\arraycolsep5pt}
   \newcommand{\bmatrix}{\arraycolsep5pt\begin{array}}
   \newcommand{\ematrix}{\arraycolsep1.5pt\end{array}}

\catcode`\@=12

\def\sgn{\mathop{\rm sgn}\nolimits}

\newtheorem{lemma}{Lemma}

\begin{document}
\title{SOLUTIONS OF THE KPI EQUATION WITH SMOOTH INITIAL DATA
\thanks{Work supported in part by Ministero delle Universit\`a e
della Ricerca Scientifica e Tecnologica, Italy}}
\author{M.~Boiti \and
F.~Pempinelli \and A.~Pogrebkov\thanks{Permanent address:  Steklov
Mathematical Institute, Vavilov Str. 42, Moscow 117966, GSP-1, RUSSIA;
e-mail POGREB@QFT.MIAN.SU}\\ Dipartimento di Fisica dell'Universit\`a e
Sezione INFN\thanks{e-mail: BOITI@LECCE.INFN.IT and
PEMPI@LECCE.INFN.IT}\\ 73100 Lecce, ITALIA}
\date{June 23, 1993}
\maketitle

\begin{abstract}
The solution $u(t,x,y)$ of the Kadomtsev--Petviashvili I (KPI) equation
with given initial data $u(0,x,y)$ belonging to the Schwartz space is
considered. No additional special constraints, usually considered in
literature, as $\int\!dx\,u(0,x,y)=0$ are required to be satisfied
by the initial data. The problem is completely solved in the framework
of the spectral transform theory and it is shown that $u(t,x,y)$
satisfies a special evolution version of the KPI equation and that, in
general, $\partial_t u(t,x,y)$ has different left and right limits at
the initial time $t=0$. The conditions of the type $\int\!dx\,u(t,x,y)=0$,
$\int\!dx\,xu_y(t,x,y)=0$ and so on (first, second, etc. `constraints')
are dynamically generated by the evolution equation for $t\not=0$.  On the
other side $\int\!dx\!\!\int\!dy\,u(t,x,y)$ with prescribed order of
integrations is not necessarily equal to zero and gives a nontrivial
integral of motion.
\end{abstract}


\section{Introduction}

We consider the Kadomtsev--Petviashvili
equation~\cite{kp,brown,ablseg} in its version called KPI
\beq
\label{kp1}
(u_{t}-6uu_{x}+u_{xxx})_{x}=3u_{yy},\qquad u=u(t,x,y),
\eeq
for $u(t,x,y)$ real. Already in 1974~\cite{dryuma} it has been
acknowledged to be integrable since it can be associated to a linear
spectral problem and, precisely, to the non--stationary Schr\"odinger
equation
\beq
\label{Schrodin}
(-i\partial _{y}+\partial ^{2}_{x}-u(x,y))\Phi =0.
\eeq
However, to building a complete and coherent theory for
the spectral transform of the potential $u(x,y)$ in (\ref{Schrodin}) that
could be used to linearize the initial value problem of (\ref {kp1})
resulted to be unexpectedly difficult. The real breakthrough has been the
discover that the problem is solvable via a non local Riemann--Hilbert
formulation~\cite{ZManakov,FokasAblowitz}.

Successively other progresses have been made. The characterization
problem for the spectral data was solved in~\cite{DS+KP}.  The extension of
the spectral transform to the case of potential $u(x,y)$ approaching to
zero in every direction except a finite number has been given
in~\cite{total}. The questions of the associated conditions
(often called `constraints') and how to
choose properly $\partial^{-1}_x$ in the evolution form of KPI
\beq
\label{kp1evol}
u_{t}(t,x,y)-6u(t,x,y)u_{x}(t,x,y)+u_{xxx}(t,x,y)=3\partial^{-1}_x\,
u_{yy}(t,x,y)
\eeq
have been studied in~\cite{ablvill}. Some additional relevant points on
constraints are known, but not yet published~\cite{Manakov}.

In our opinion, the problem of the proper choice of $\partial^{-1}_x$ in
(\ref{kp1evol}) requires an additional investigation paying special
attention to the behavior of $u$ and spectral data at the initial time,
say at $t=0$. This study can now be done since we have at hands some
additional theoretical tools in the theory of the spectral transform
developed in~\cite{total,resreg,nsreshi}.

In the following we will find convenient to consider the problem also in
terms of the Fourier transform of $u$
\beq
\label{fourier}
v(t,p)\equiv v(t,p_1,p_2)={1\over (2\pi)^{2}}\int\!\!\!\int\!\!dxdy\,
e^{i(p_1x-p_2y)}u(t,x,y),
\eeq
where $p$ denotes a 2--component vector
\beq
\label{2comp}
p\equiv \{p_1,p_2\}.
\eeq
Then, the evolution equation (\ref {kp1evol}) can be
rewritten as $(d^2q\equiv dq_1dq_2)$
\beq
\label{kp1evol'} {\partial v(t,p)\over \partial
t}=-3ip_{1} \!\int\!\!d^{2}q\,v(t,q) \,v(t,p-q)-
i\,{p_{1}^{4}+3p_{2}^{2}\over p_{1}}\,v(t,p)
\eeq
and we are faced with the problem of properly defining the distribution
$p_{1}^{-1}$. The integrations here and in
the following when it is not differently indicated are performed all
along the real axis from $-\infty$ to $+\infty$.

According to the usual scheme of the spectral transform theory we expect
that the spectral data evolve in time as the linear part of
(\ref{kp1evol'}). This makes clear that the special behavior of the
considered quantities at the initial time $t=0$ is just complicated by
the nonlinearity of the evolution equation but it is, in fact, inherent to
the singular character of its linear part.

The final answer is that, for initial data $u(0,x,y)$ belonging to the
Schwartz space which are arbitrarily chosen and, therefore, not
necessarily subjected to the constraint \beq \label{t0constraint}
\int\!dx\,u(0,x,y)=0,\qquad \mbox{or}\quad v(0,0,p_2)=0,
\eeq
the function $u(t,x,y)$ reconstructed by solving the inverse spectral
problem for  (\ref{Schrodin}) evolves in time  according to the equation
\beq
\label{evolutionkp1}
u_t(t,x,y)-6u(t,x,y)u_x(t,x,y)+u_{xxx}(t,x,y)=3\!\int^{x}_{-t\infty}
\!\!dx'\,u_{yy}(t,x',y).
\eeq
Therefore, for smooth initial data not satisfying constraint
(\ref{t0constraint}), $u_t(t,x,y)$ has at $t=0$ different left and right
limits and the condition
\beq \label{constraint1}
\int\!dx\,u(t,x,y)=0,
\eeq
is dynamically generated by the evolution equation at times
$t\not=0$, and thus for these times we recover the result obtained
in~\cite{ablvill}. On the other side, the initial time $t=0$
requires a special investigation and the evolution form of the KPI equation
we deduce in (\ref {evolutionkp1}) is different from that proposed
in~\cite{ablvill}.

In the same way higher conditions (`constraints') can be obtained.
But some special care is needed as, e.g., for the next condition we have
\beq
\label{constraint2}
\int\!dx\,x\,u_y(t,x,y)=0, \qquad t\not=0,
\eeq
and the $y$-derivative cannot be extracted from the integral. In
spite of all these facts there exists
\beq \label{int}
\int\!dx\!\int\!dy\,u(t,x,y)=\int\!\int\!\!dxdy\,u(0,x,y),
\eeq
where the order of integrations in the {l.h.s.} is explicitly shown. It
is clear that this quantity is not necessarily equal to zero and gives
a nontrivial integral of motion.

In terms of the Fourier transform $v(t,p)$ we have
\beq
{\partial v(t,p)\over \partial t}=-3ip_{1} \!\int\!\!d^{2}q\,v(t,q)
\,v(t,p-q)- i\,{p_{1}^{4}+3p_{2}^{2}\over p_{1}+i0t}\,v(t,p)
\eeq
and
\beq
\lim_{p_1\to 0}v(t,p_1,p_2)=0,\qquad \mbox{for}\quad t\not=0,
\eeq
where the limit is understood in the sense of distributions in
the $p_2$ variable.


\section{The Linearized Equation}

According to the remark made in the introduction, insofar as initial value
problem
\beq
\label{lin1}
\partial_{x} (\partial_{t} U(t,x,y)+\partial^{3}_{x} U(t,x,y))=
3\partial^{2}_{y}U(t,x,y),
\eeq
\beq
\label{init}
U(0,x,y)=U(x,y)
\eeq
is concerned, we expect the linear equation
and the nonlinear one (\ref {kp1}) to be closely related. We consider here
the case of initial data $U(x,y)$ belonging to the test function Schwartz
space ${\cal S}$ in the $x$ and $y$ variables.


The study of the existence and uniqueness of the solution of the initial
value problem defined in (\ref {lin1}), (\ref {init}) requires to pay special
attention to the behavior of the solution $U(t,x,y)$ at $t=0$. In fact if
this function is continuously differentiable proper number of times in
$x$, $y$ and $t$ variables
and vanishing with its derivatives for $(x,y)\to \infty$ at any
time, then the initial value problem has no solution if the initial data does
not satisfy the constraint (\ref {t0constraint}). However we show in the
following that if  for $t=0$ (and only for $t=0$) $U_t(t,x,y)$ is
allowed to be discontinuous and not decreasing at large ($x,y$) then
the solution $U(t,x,y)$ of the initial value problem exists and is
uniquely determined for any initial data $U(x,y)\in {\cal S}$.  In
addition an evolution form of (\ref {lin1}) satisfied by $U(t,x,y)$
at all times including $t=0$ (in the sense of left/right limits) is
uniquely determined.

First of all we see that if such $U(t,x,y)$ exists then, by
integrating both sides of (\ref {lin1}) for any $t\neq0$, we can
rewrite (\ref {lin1}) in the form
\beq \label{linevol1} \partial
_{t}U(t,x,y)=-\partial ^{3}_{x}U(t,x,y)+3
\int^{x}_{-\infty}dx''\,U_{yy}(t,x'',y),\quad t\neq 0,
\eeq
and derive, again for any   $t\neq 0$, the condition
\beq
\label{cond1'}
\!\int\!\!dx\,U_{yy}(t,x,y)=0.
\eeq
Let $V(t,p)$ and $V(p)$ denote the Fourier transforms of
$U(t,x,y)$ and $U(x,y)$ according to the definition (\ref {fourier}). Then
(\ref {linevol1}) gives
\beq
\label{linevol2}
{\partial V(t,p)\over \partial t}=-i\,{p_{1}^{4}+3p_{2}^{2}\over
p_{1}}\,V(t,p)\qquad t\neq 0
\eeq
and due to (\ref {init})
\beq\label{init2}
V(0,p)=V(p)\in{\cal S}.
\eeq
Notice that the exact meaning of $p_1^{-1}$ in (\ref {linevol2}) is
irrelevant due to condition~(\ref {cond1'}). For any $p_1\neq 0$ the
unique solution of the problem~(\ref {linevol2}),~(\ref {init2}) is given as
\beq
\label{1}
V(t,p)=\exp\left(-it\,{p_{1}^{4}+3p_{2}^{2}\over p_{1}}\right)\,V(p).
\eeq

To proceed further we need the following
\begin{lemma}\label{K1}
Function $\exp(i\tau / p_1)$ defines a distribution, depending continuously on
parameter $\tau $, in the Schwartz space of the variable $p_1$.
\begin{enumerate}
\item[{\sl i)}] This distribution is continuously differentiable
in $\tau $ for any $\tau \neq 0:$
\begin{equation}
\label{deriv1}
\partial_{\tau }\exp\left({i\tau \over p_1}\right)=
{i\over p_1} \,\exp\left({i\tau \over p_1}\right),\qquad \tau \neq 0,
\end{equation}
where the {r.h.s.} is a well defined distribution in the same space.
\item[{\sl ii)}] At $\tau =0$ there exist right/left limits
\begin{equation}
\lim_{\tau \to \pm 0}{1\over p_{1}}\,
\exp\left({i\tau \over p_1}\right)= {1\over p_1\mp i0}.\nonumber
\end{equation}
\end{enumerate}
Thus due to (\ref {deriv1})  we can write for arbitrary $\tau $
\begin{equation}
\label{derivexp}
\partial_{\tau }\exp\left({i\tau \over p_1}\right)=
{i\over p_1-i0\tau } \,\exp\left({i\tau \over p_1}\right).
\end{equation}
\end{lemma}
{\sl Proof}

{\sl i)} Let us consider
\begin{equation}
f(\tau )=\!\int\!\! dp_1\,e^{i\tau/p_1}
\varphi (p_1), \end{equation}\nonumber
where $\varphi (p_1)\in {\cal S}$. Subtracting and adding terms we can write
\beqsn
f(\tau )=\!\int\!\!dp_1\,e^{i\tau/p_1}
\bigl[ \varphi (p_1)-\vartheta (1-\mid p_1\mid ) \varphi (0) \bigr]
+\varphi (0)\!\int^{1}_{-1}\!\!dp_1 \,e^{i\tau/p_1}.
\eeqsn
Changing $p_1$ to $1/p_1$ in the second term we have
\beqs
\label{deff}
f(\tau)=\!\int\!\!dp_1\,e^{i\tau/p_1} \bigl[
\varphi (p_1)-\vartheta (1-\mid p_1\mid ) \varphi (0) \bigr] +
\varphi (0)\!\int_{|p_1|\geq1}\!\!{dp_1\over p_1^2}  \,e^{i \tau
p_1 }.
\eeqs
Now it is obvious that both terms are differentiable in $\tau $
for $\tau \neq 0$ and
\beqs
\label{derf}
\partial_{\tau }f(\tau )=i\!\int\!\!{dp_1 \over p_1}\,
e^{i\tau/p_1} \bigl[ \varphi (p_1)-\vartheta (1-\mid
p_1\mid ) \varphi (0) \bigr] +
i\varphi (0)\!\int_{|p_1|\geq1}\!\!{dp_1\over p_1}\,
e^{i\tau p_1 } .
\eeqs
Substituting again $p_1$ for $1/p_1$ in the second term we obtain
that it cancels out with the second term in brackets.  This
proves {\sl i)}. Let us remark that (\ref {derf}) shows that for
$\tau \neq 0$ it is  not necessary to regularize the factor $1/p_1$ in
front of the exponent in (\ref {deriv1}).   One can consider this
factor indifferently as principal value or as $(p_1\pm i0)^{-1}$.

{\sl ii)} Again subtracting and adding terms we can write
\beqsn
\!\int\!\! {dp_1\over p_1} \,e^{i\tau/p_1}\,
\varphi (p_1) &=&
\int  {dp_1\over p_1} e^{i\tau/p_1} \left[
\varphi (p_1)-\vartheta (1-\mid p_1\mid ) \varphi (0) \right] +\\
&&\sgn \tau \,\varphi (0)\int_{|p_1|\geq|\tau |} {dp_1\over p_1} \,e^{i
p_1} ,
\eeqsn
where now in the second term $p_1$ was substituted for $\tau p_1$.
To compute the limit for $\tau \to 0$ notice that
\begin{equation}
\lim_{\tau \rightarrow  0}\int_{\mid p_1\mid\geq\mid \tau \mid }
{dp_1\over p_{1}}\,e^{ip_1} =
i\pi.
\end{equation}
Thus
\beqsn
\lim_{\tau \rightarrow  0}\int {dp_1\over p_1}\,
e^{i\tau/p_1} \varphi (p_1)& =&
\int  {dp_1\over p_1} \left[ \varphi (p_1)-\vartheta
(1-\mid p_1\mid ) \varphi (0) \right] \pm i\pi
\varphi (0)\equiv\\
&&\int  {dp_1\over p_1} \, \varphi (p_1)\pm i\pi\varphi (0),
\eeqsn
where to get the second equality we chose $p_1^{-1}$ in the
integral as the principal value. Then (\ref {derivexp}) follows due
to (\ref {deriv1}) and the remark made after the proof of
{\sl i)}.  $\Box$

Applying this Lemma we see that
$V(t,p)$ given in (\ref {1}), for $V(p)$ belonging  to the
test--function space $\cal S$ of Schwartz, is a distribution (in the
$p_1$ and $p_2$ variables) belonging to the dual space $\cal S'$
and that $V(t,p)$ obeys for any $t$ the equation
\beq
\label{linevol3} {\partial V(t,p)\over \partial
t}=-i\,{p_{1}^{4}+3p_{2}^{2}\over p_{1}+i0t}\,V(t,p).
\eeq
Note that the $i0$-term in the denominator is relevant only for $t\to
\pm 0$ since for any $t\not=0$ due to rapid oscillations in (\ref
{1})
\beq \label{cond1''} \lim_{p_1\to 0}V(t,p_1,p_2)=0
\eeq
in the sense of distributions in $p_2$. Thus $U_t$ is discontinuous at
$t=0$.

These properties of the distribution
\beq
\exp\left(-it\,{p_{1}^{4}+3p_{2}^{2}\over p_{1}}\right)
\eeq
can be received as well by noting that it can be obtained as a limit in
the sense of ${\cal S}'$  according to the following formula
\beq
\exp\left(-it\,{p_{1}^{4}+3p_{2}^{2}\over p_{1}}\right)
=\lim_{\epsilon\to +0} \exp\left(-it\,{p_{1}^{4}+3p_{2}^{2}\over
p_{1}+i\epsilon t}\right).
\eeq
Then, since in the {r.h.s.} we have functions belonging to ${\cal S}$
for $t\neq 0$, we can use the property that the derivation with respect
to $t$ is a continuous operation in the space of distributions in the
variables $t$, $p_1$ and $p_2$ (cf.  \cite{gelfsh}).

Coming back to $U(t,x,y)$ by making the inverse Fourier transform
\beq
\label{Ut}
U(t,x,y)=\!\int\!\!\!\int\!\!d^{2}q\, e^{- i q_{1}x+i q_{2}y-i t q^{3}_{1}-3i
t {q ^{2}_{2}\over q_{1}} } V(q)
\eeq
we get that it exists and solves the initial value problem (\ref
{lin1}), (\ref {init}). To study the properties of this solution it is
convenient  to extract
the nonsingular part of the dispersive function in the exponent, i.e.
the part proportional to $q_1^3$.  We introduce
\begin{equation}\label{defU0} U_{0}(t,x,y) = \!\int\!\!
d^{2}q\,e^{-iq_{1}x+iq_{2}y-iq^{3}_1t}\, V(q), \end{equation}
which
solves the equation
\begin{equation}\label{eqU0} \partial
_{t}U_{0}(t,x,y)=-\partial ^{3}_{x}U_{0}(t,x,y).
\end{equation}
and belongs to ${\cal S}$ in the $x$ and $y$ variables for any $t$
since $V(p)\in {\cal S}$.
Then from (\ref {Ut}) we have
\begin{equation}
\label{UtJ} U(t,x,y) =
\!\int\!\!\!\int\!\!  dx'dy'\,J(t,x-x',y-y')\,U_{0}(t,x',y'),
\end{equation}
where we have introduced the distribution
\begin{equation} \label{defJ} J(t,x,y)=\!\int\!\!\!\int\!\!d^2 p\,
\exp\left(-ixp_1+iyp_2-3it\,{p_{2}^{2}\over p_{1}}\right).
\end{equation}
By computing the Fourier transform we get
\begin{equation} \label{explJ}
J(t,x,y) = {\sgn t\over \pi } \partial_{x}(12xt-y^{2})^{-1/2}_+ ,
\end{equation}
where the standard definition (cf.~\cite{gelfsh})
\begin{equation}\label{notation}
x_+^{-1/2}=\vartheta (x)\frac{1}{\sqrt{x}}
\end{equation}
is used. The distribution $J(t,x,y)$ obeys the following properties
\begin{equation}\label{initJ}
J(0,x,y) = \delta (x)\,\delta (y),
\end{equation}
\begin{equation}\label{derJ}
\partial _{t}J(t,x,y) = 3\partial ^{2}_y\!\int^{x}
_{-t\infty}\!\!dx'\,J(t,x',y ),
\end{equation}
\begin{equation}\label{derJpm}
\lim_{t\rightarrow \pm 0} \partial_tJ(t,x,y) =
\pm 3\vartheta (\pm x)\,\partial^2_y\delta(y).
\end{equation}
Now by means of (\ref {UtJ}) and (\ref {explJ}) it is easy to check
that (\ref {Ut}) gives the solution of (\ref{lin1})
obeying all properties mentioned above. To prove that this solution is unique
let us notice that any solution of the problem (\ref {linevol2}),~(\ref
{init2}) can differ from the one given in (\ref {1}) only for a
distribution localized at $p_1=0$, i.e. for terms proportional to $\delta
(p_1)$ or its derivatives. Addition of such terms in the r.h.s. of the
(\ref {Ut}) violates the condition of decreasing of $U(t,x,y)$ for large
$x$. Thus in the chosen class the solution (\ref {Ut}) is unique.

Recall that we imposed to the solution to be
continuously differentiable in $t$ only for $t\neq 0$. We have
therefore to study separately the properties of
(\ref {Ut}) at $t=0$.  Due to (\ref {UtJ}), (\ref {eqU0}), (\ref
{derJ}) and (\ref {derJpm}) we derive that $U(t,x,y)$ satisfies the
evolution equation
  \beq \label{linevol4} \partial
_{t}U(t,x,y)=-\partial ^{3}_{x}U(t,x,y)+3\partial ^{2}_y
\int^{x}_{-t\infty}dx'U(t,x',y), \eeq
and that the condition
\beq
\label{constr}
\!\int\!\!dx\,U(t,x,y)=0,\qquad t\neq 0,
\eeq
is generated dynamically. Of course (\ref {linevol4}) and (\ref
{constr}) are just the Fourier transforms of (\ref {linevol3}) and (\ref
{cond1''}).

In spite of (\ref {constr}) integrating (\ref {UtJ}) first in $y$ and
then in $x$ we get due to (\ref {defJ}) and (\ref {explJ}) that
\beq
\label{integrallin}
\!\int\!\!dx\!\int\!\!dy\,U(t,x,y)=V(0)=\!\int\!\!\!\int\!\!dxdy\,U(x,y)
\eeq
for any $t$, i.e. for this order of integrations the {l.h.s.} gives an
integral of motion, which is not equal to zero in a generic situation.

By means of (\ref {UtJ}), (\ref {defJ}) and (\ref {explJ})  we  derive
that  $U(t,x,y)$, for $t$ ($t\neq 0$) and $y$ fixed, decreases rapidly
for $x\to -t\infty$ while for $x\to t\infty$ it
satisfies the following asymptotic behavior
\beqs
\label{asymptotics}
U(t,x,y)&= &{-1\over 4\pi \sqrt{3tx}|x|}\!\int\!\!\!\int\!\!dx'dy'\,U(x',y')\\
& &-{1\over 32\pi |t|\sqrt{3tx}x^2}\!\int\!\!\!\int\!\!dx'dy'
\,[12tx'+(y-y')^2]\,U(x',y')+o(|x|^{-5/2}).\nonumber
\eeqs
This asymptotic expansion is differentiable it $t$ and twice differentiable
in $y$ and obeys (\ref {lin1}). Therefore we can differentiate in $t$
the condition~(\ref {constr}) and use (\ref {linevol4}). This procedure
leads to the next condition
\beq \label{constr2'}
\!\int\!\!dx\,x\,U_{yy}(t,x,y)=0,\qquad t\neq 0.
\eeq
Thanks to  the
asymptotic behaviour of $U(t,x,y)$ at large $x$ derived in (\ref
{asymptotics}) one $y$-derivative can be extracted from the
integral getting
\beq \label{constr2}
\!\int\!\!dx\,x\,U_{y}(t,x,y)=0,\qquad t\neq 0,
\eeq
but it is impossible to remove the second derivative since
$\int\!\!dx\,x\,U(t,x,y)$ is divergent. This procedure can be continued to
get an infinite set of dynamically generated conditions.

Equation (\ref {asymptotics}) explains the role of constraints,
i.e. conditions of the type~(\ref {constr}) and~(\ref {constr2})
imposed on the initial data $U(x,y)$, in the asymptotic behaviour of
$U(t,x,y)$ at large $x$. Precisely, for each additional constraint that
is satisfied we get an additional $x^{-1}$ factor in the decreasing
law.

Finally we notice that it is possible to weaken the conditions on the
initial data. For example, it is enough for $V(p_1,p_2)$ to be
continuous at point $p=0$ separately in $p_1$ and $p_2$.

In the nonlinear case (\ref {kp1}) we will consider solutions $u(t,x,y)$
that have the same mentioned smoothness
properties in the $x$, $y$ and $t$ variables as $U(t,x,y)$. The
possible discontinuity of $u_t(t,x,y)$ at $t=0$ can be factorized by
considering the following representation for the Fourier transform
$v(t,p)$ of $u(t,x,y)$
\beq
\label{tilde} v(t,p)=\exp\left(-it\,{p_{1}^{4}+3p_{2}^{2}\over
p_{1}}\right) \,\widetilde{v}(t,p),
\eeq
which can be considered a natural generalization of (\ref{1}).
Then $\widetilde{v}(t,p)$ is required to be continuously differentiable
also at $t=0$ and continuous at the point $p=0$ separately in $p_1$ and
$p_2$.

\section{The Nonlinear Equation}

\subsection{The direct problem at $t=0$}
The direct problem has been extensively studied
in~\cite{ZManakov,FokasAblowitz,total,resreg,nsreshi}. We report here
the main definitions and formulae we need in the following
emphasizing specific features of the considered problem.

The Jost solutions $\Phi(x,y|k)$  are special eigenfunctions of the
non--stationary Schr\"o\-din\-ger equation (\ref{Schrodin}), which are
analytic
in the upper and lower half plane of the complex spectral parameter $k$ and
whose values $\Phi^{\pm}(x,y|k)$ on the two sides $\pm\Im k>0$ of the real
$k$--axis
\beq \label{phimu}
\Phi ^\pm (x,y|k) = e^{-ikx+ik^2y}\,\mu ^\pm
(x,y|k) \eeq
are determined by the integral equation
\beq \label{dirJost}
\mu ^\pm
(x,y|k)=1+\!\int\!\!\!\int\!\!dx'dy'\,G^{\pm}(x-x',y-y'|k)\,u(x',y') \,\mu
^\pm (x',y'|k),
\eeq
where
\beq \label{GreenJost} G^{\pm}(x,y|k)={1\over
(2\pi)^2}\!\int\!\!d^2p\,{e^{-ip_1x+ip_2y}\over {\cal L}(p)-2p_1k\mp i0p_1}
\eeq
and
\beq
\label{L}
{\cal L}(p)=p_2-p_1^2.
\eeq
The special role played by the spectral parameter $k$ is stressed by
separating it from other variables by a vertical bar.
The Green function introduced in (\ref {GreenJost}) obviously can be
analytically continued in the upper and bottom half planes in $k$:
\beq
\label{GreenJost'}
G^{\pm}(x,y|k)={1\over
(2\pi)^2}\!\int\!\!d^2p\,{e^{-ip_1x+ip_2y}\over {\cal L}(p)-2p_1k},
\qquad \pm\Im k>0.
\eeq
It decreases for $k\to\infty$ and in what follows we need the first
coefficient of its $1/k$ expansion. By means of (\ref {GreenJost'}) we
have
\beqs \label{GreenJost1}
&&\lim_{k\to\infty}kG^{\pm}(x,y|k)=\lim_{k\to\infty}{-1\over
2(2\pi)^2}\!\int\!\!d^2p\, {e^{-ip_1x+ip_2y}\over p_1-{{\cal L}(p)\over
2k} }=\\ &&{-1\over 2(2\pi)^2}\!\int\!\!d^2p\,
{e^{-ip_1x+ip_2y}\over p_1\pm i0p_2}\equiv {1\over 4\pi}\left({\mp 1\over y}
+i\pi\sgn x\,\delta (y)\right), \qquad \pm\Im k>0.\nonumber
\eeqs
Thus we see that the term in the expansion which is nonlocal in $y$
explicitly depends on the half plane of $k$ in which the limit is
performed.  Correspondingly for the first two terms in the $1/k$
expansion of $\mu$ we have
\beqs \label{mu1}
\mu^{\pm} (x,y|k)=1&+&
{1\over 4\pi k}\left(\mp\!\int\!\!\!\int\!\!{dx'dy'\over
y-y'}\,u(x',y')+
i\pi\!\int\!\!dx'\,\sgn(x-x')\,u(x',y)\right)+\nonumber\\
&&o\left({1\over k} \right),\quad k\to\infty,\quad \pm\Im k>0,
\eeqs
which also demonstrates the dependence on the half plane.

The spectral data are defined as
\beq
\label{SD}
r^{\pm}(\alpha ,\beta)={1\over (2\pi)^2}\!\int\!\!\!\int\!\!dxdy\,
e^{ix(\alpha -\beta)-iy(\alpha ^2-\beta ^2)}\,u(x,y)\,\mu^{\pm}(x,y|\beta).
\eeq
Note that, due to specific dependence of the exponent in (\ref
{SD}) on $\alpha $ and $\beta $, spectral data decrease for large values of
the arguments only if $\alpha -\beta $ or $\alpha^2
-\beta^2$ increase. It means that if we consider for some fixed parameter
$\gamma$
\beq
r^{\pm}\left(\beta +{\gamma \over \beta
},\beta  \right) ={1\over (2\pi)^2}\!\int\!\!\!\int\!\!dxdy\,
e^{ix{\gamma \over \beta} -
iy\left(2\gamma +{\gamma ^2\over \beta ^2} \right)}\,u(x,y)\,
\mu^{\pm}(x,y|\beta),  \nonumber
\eeq
then, since $\mu(x,y|\beta )\to 1$ for $\beta \to\infty$,
\beq
\label{lim1}
\lim_{\beta \to\infty} r^{\pm}\left(\beta +{\gamma \over \beta },\beta
\right)
={1\over (2\pi)^2}\!\int\!\!\!\int\!\!dxdy\,
e^{-2iy\gamma}\,u(x,y),
\eeq
which is not necessarily equal to zero. This remark will be crucial in
the following.

We need in the following to consider the counterpart of all these
formulae in the Fourier transformed space. The Fourier transformed Jost
solution is defined as
\beq \label{munuJ}
\nu^\pm
(p|k)=\!\int\!\!dx\,dy\,e^{ip_1x-ip_2y}\,\mu^\pm (x,y|k), \qquad \pm\Im
k>0, \eeq
and satisfies the integral equation
\beq
\label{dirJost'}
\nu^\pm (p|k)=\delta^2 (p)+{1\over {\cal L}(p)-2p_1k\mp i0p_1}\!\int\!\!d^2p'
\,v(p-p') \,\nu^\pm(p'|k),
\eeq
where notation~(\ref {L}) has been used and $\delta
^2(p)=\delta(p_1)\delta (p_2)$. From this equation the expansion in
$1/k$ can be easily derived. For following use we write it up to the third
term:
\beqs
\label{nuJost2}
\nu^\pm (p|k)&=&\delta^2 (p)-{1\over 2k}{v(p)\over p_1\pm i0p_2} -\\
& &{1\over
(2k)^2}\left\{{{\cal L}(p)\,v(p)\over (p_1\pm i0p_2)^2}-
{1\over p_1\pm i0p_2} \!\int\!\!d^2p'\, {v(p-p')\,v(p')\over p'_1\pm i0p'_2}
\right\} +\nonumber\\
&&o\left({1\over k^2} \right),\qquad
\pm\Im k>0, \qquad k\to\infty.\nonumber
\eeqs
Formula (\ref{lim1}) reads
\beq
\label{lim1'}
\lim_{\beta \to\infty} r^{\pm}\left(\beta +{\gamma \over \beta },\beta
\right)
=v(0,2\gamma).
\eeq
If we introduce the function
\beq
\label{rhoJost}
\rho^\pm (p|k)=[{\cal L}(p)-2p_1k]\,\nu^\pm (p|k),
\eeq
the spectral data~(\ref {SD}) can be written as
\beq
\label{SD1}
r^{\pm}(\alpha ,\beta )=\rho ^{\pm}(\ell(\alpha )-\ell(\beta )|\beta ),
\eeq
or
\beq
\label{SD2}
r^{\pm}(\alpha ,\beta )=\!\int\!\!d^2q\,v(q)\,\nu^{\pm}(\ell(\alpha
)-\ell(\beta )-q|\beta ),
\eeq
where the special two component vector
\beq
\label{l}
\ell(k)=(k,k^2)
\eeq
has been introduced. Note that for $u$ real (as we are considering)
$r^\sigma(\alpha,\beta)=\overline{r^{-\sigma}(\beta,\alpha)}$.

For more details the reader can refer to~\cite{nsreshi}, where a slightly
different notation has been used. Precisely one has to make the following
identifications
\beqs
&&\nu^{\pm}(p|k)\equiv\nu^{\pm}(p,\ell(k))\\
&&\rho ^{\pm}(\ell(\alpha )-\ell(\beta )|\beta )\equiv
\rho^{\pm}(\ell(\alpha )-\ell(\beta ),\ell(\beta)).
\eeqs

\subsection{The inverse problem at $t=0$}

The inverse problem can be solved equivalently in the $(x,y)$ space or
in the Fourier transformed $(p_1,p_2)$ space. However, we prefer to work in
the $p$-space because, as we will see in the following section, the time
evolution can be more easily handled.

We start from the known fact that the advanced/retarded eigenfunctions
$\nu_{\pm}(p|k)$ of the non--stationary Schr\"o\-din\-ger equation can be
indifferently related to the Jost solution $\nu^\sigma(p|k)$ for $\sigma=+$
or for $\sigma=-$ according to the formula
\beq
\label{sigmaind}
\nu_{\pm}(p|k)=\nu ^{\sigma }(p|k)\mp 2i\pi\!\int\!\!d\beta \, \theta
(\pm\sigma (\beta -k))\, \overline{r^{-\sigma }(k,\beta )}\,\nu ^{\sigma
}(p+\ell(k) -\ell(\beta )|\beta),\quad \sigma=+,-,
\eeq
i.e. the {l.h.s.} is independent on the sign of $\sigma $
(for a proof see~\cite{total,resreg,nsreshi,chinese}). From
(\ref{sigmaind}) we deduce that also the sum
$\nu_{+}(p|k)+\nu_{-}(p|k)$ is $\sigma$ independent or that
\beqs
\label{sigmaind1}
&&\nu^{\sigma }(p|k)+
i\pi\sigma \!\int\!\!d\beta \, \sgn (k-\beta)\, \overline{r^{-\sigma
}(k,\beta )}\,\nu ^{\sigma }(p+\ell(k) -\ell(\beta )|\beta)=\\&&(\sigma
\to-\sigma ).\nonumber
\eeqs
  Due to this the non--local Riemann--Hilbert
problem for the discontinuity of the Jost solution across the real
k--axis (see also \cite{nsreshi}) is given by the equation
\beq
\label{discJost}
\nu^{+}(p|k)-\nu ^{-}(p|k)=-
i\pi\!\!\sum_{\sigma =+,-}\!\int\!\!d\beta \, \sgn (k-\beta)\,
\overline{r^{-\sigma }(k,\beta )}\,\nu ^{\sigma} (p+\ell(k) -\ell(\beta
)|\beta),
\eeq
that has to be compared with the usual one in literature
which is quadratic in the spectral data.

{}From the analytic properties of $\nu ^\pm$ and the Cauchy formula we
have that  $\nu ^\pm$ can be reconstructed from the spectral data
$r^{\pm}(k,\beta )$ by solving the singular linear integral equation
\beq
\label{IPJost2}
\nu^\pm (p|k)=\delta ^2(p)-\!\int\!\!{d\alpha
\over \alpha -k\mp i0} \sum_{\sigma =+,-}\!\int\!\!d\beta \,{\sgn (\alpha
-\beta )\over 2} \, \overline{r^{-\sigma }(\alpha ,\beta )}\,\nu^{\sigma
}(p+\ell(\alpha )-\ell(\beta )|\beta ).
\eeq
One can use the explicit $k$--dependence in the {r.h.s.} to get a $1/k$
expansion at large $k$ of $\nu^\pm (p,k)$. However, before expanding the
denominator in the r.h.s. it is necessary to perform a subtraction in
the integration over $\alpha$. In fact, due to (\ref {lim1'}) and
(\ref{nuJost2}), we have (for any sign of $\alpha$)
\beqs
\label{lim6}
&&\lim_{\alpha
\to\infty}\alpha \!\int\!\!d\beta \,{\sgn (\alpha -\beta )\over 2} \,
\overline{r^{-\sigma }(\alpha ,\beta )}\,\nu^{\sigma }(p+\ell(\alpha )-
\ell(\beta ),\beta )=\\
&&\lim_{\alpha\to\infty}{1\over 2} \!\int\!\!d\beta \,\sgn \beta \,
\overline{r^{-\sigma }\left(\alpha ,\alpha -{\beta \over\alpha}\right)}\,
\nu^{\sigma }\left(p_1+{\beta \over \alpha },p_2+{\beta \over \alpha }
\left(2\alpha -{\beta \over \alpha }\right)\left|\alpha -{\beta
\over \alpha }\right. \right)=\nonumber\\ &&-{\delta (p_1)\over
4}\,\sgn p_2 \,v(p),\nonumber \eeqs where to get the second line the
substitution $\beta \to \alpha -\beta/ \alpha$ has been performed.
This proves that a direct expansion would furnish a divergent result.

Therefore, we rewrite (\ref {IPJost2}) as
\beqs
\label{IPJost2'}
\nu^\pm(p|k)&=&\delta ^2(p)-\\
&&\!\int\!\!{d\alpha \over
\alpha -k\mp i0}\Bigl\{\sum_{\sigma =+,-} \!\int\!\!d\beta \,
{\sgn (\alpha -\beta )\over 2} \, \overline{r^{-\sigma }(\alpha ,\beta
)}\,\nu^{\sigma }(p+\ell(\alpha ) -\ell(\beta )|\beta )+\nonumber\\
&&{\delta (p_1)\over 2(\alpha \mp i0)} \,\sgn p_2\,v(p)\Bigr\}.\nonumber
\eeqs
The additional term corrects the bad behavior of the integrand at large
$\alpha$ and does not modify the total value of the {r.h.s.} of
(\ref{IPJost2}) since, once integrated over $\alpha$, gives a zero
contribution.

Now, we can compute the first term in the expansion at large $k$ and we get
\beqs
\label{nuJost3}
&&\lim_{k\to\infty}k[\nu^\pm (p|k)-\delta ^2(p)]=\\
&&\!\int\!\!d\alpha
\Bigl\{\sum_{\sigma =+,-} \!\int\!\!d\beta \,{\sgn
(\alpha -\beta )\over 2} \, \overline{r^{-\sigma }(\alpha ,\beta
)}\,\nu^{\sigma }(p+\ell(\alpha )-\ell(\beta )|\beta )+\nonumber\\
&&\qquad\qquad\qquad{\delta (p_1)\over 2(\alpha \mp i0)
}\,\sgn p_2\,v(p)\Bigr\},\quad \pm\Im k>0.\nonumber
\eeqs
Then, by using (\ref {nuJost2}), we obtain
\beqs
\label{poten5}
&&{v(p)\over p_1\pm i0p_2}= \\
&&-\!\int\!\!d\alpha \Bigl\{\sum_{\sigma =+,-} \!\int\!\!d\beta \,\sgn (\alpha
-\beta )\, \overline{r^{-\sigma }(\alpha ,\beta)}\,\nu^{\sigma }(p+\ell(\alpha
)-\ell(\beta )|\beta ) +{\delta (p_1)\over \alpha \mp i0 }\,\sgn
p_2\,v(p)\Bigr\}\nonumber
\eeqs
or, taking into account that the delta distributions coming from the $\pm
i0$ terms cancel out, we get
\beq
\label{poten5'}
{v(p)\over p_1}= -\!\int\!\!d\alpha
\Bigl\{\sum_{\sigma =+,-} \!\int\!\!d\beta \,\sgn (\alpha -\beta )\,
\overline{r^{-\sigma }(\alpha ,\beta)}\,\nu^{\sigma }(p+\ell(\alpha
)-\ell(\beta )|\beta ) +{\delta (p_1)\over \alpha }\,\sgn p_2\,v(p)\Bigr\},
\eeq
where $1/p_1$ and $1/\alpha$ have the sense of principal value.

Finally we solve the inverse problem by reconstructing the potential $v(p)$
in terms of the spectral data and Jost solutions according to the following
formula
\beq
\label{poten6}
v(p)= -\!\int\!\!d\alpha
\,p_1\,\sum_{\sigma =+,-} \!\int\!\!d\beta \,\sgn (\alpha -\beta )\,
\overline{r^{-\sigma }(\alpha ,\beta)}\,\nu^{\sigma }(p+\ell(\alpha
)-\ell(\beta )|\beta ),
\eeq
where we used that $p_1\delta (p_1)=0$. Note that due to this fact
$p_1$ cannot be extracted from the integral.


\subsection{The time evolution}

The time evolution in (\ref {IPJost2}) and (\ref {poten6}) is switched on
by choosing spectral data that depend parametrically on the time as
follows
\beqs
r^\sigma(t,\alpha,\beta)=e^{-4it(\alpha^3-\beta^3)}r^\sigma(\alpha,\beta).
\eeqs
We need, then, to compute directly from
\beqs
\label{IPJost''}
&&\nu^\pm(t,p|k)=\delta ^2(p)-\\
&&\!\int\!\!{d\alpha \over \alpha -k\mp i0}\,\sum_{\sigma =+,-}
\!\int\!\!d\beta\,e^{4it(\alpha ^3-\beta ^3)} \, {\sgn (\alpha -\beta )\over
2} \, \overline{r^{-\sigma }(\alpha ,\beta)}\, \nu^{\sigma }
(t,p+\ell(\alpha )-\ell(\beta )|\beta )
\nonumber
\eeqs
the evolution equation satisfied by the Jost solution $\nu^\pm(t,p|k)$, i.e.
the second equation of the Lax pair. We expect it to have the form of the
Fourier transform of the standard one
\beqs
\label{dtnunot0}
{\partial\over \partial t} \nu^{\sigma}(t,p|k)&= &-2i[3(p_1+k){\cal
L}(p)+2p_1^3]\, \nu^{\sigma}(t,p|k)- \\ & &3i\!\int\!\!\!\int\!\!d^2p'\,{\cal
L}(p')\,{v(t,p')\over p'_1} \, \nu^{\sigma}(t,p-p'|k),\qquad
t\not=0,\nonumber
\eeqs
where possibly the sense of $1/p'_1$ must be specified. Then we have to prove
that
\beq
\label{poten:tJ}
v(t,p)=-\!\int\!\!d\alpha\,p_1\,\sum_{\sigma =+,-} \sigma
\!\int\!\!d\beta\,e^{4it(\alpha ^3-\beta ^3)} \,
\sgn (\alpha -\beta )\, \overline{r^{-\sigma }(\alpha ,\beta)}\,
\nu^{\sigma }(t,p+\ell(\alpha )-\ell(\beta )|\beta )
\eeq
satisfies the Fourier transform of the KPI equation.
Note that for $t\not=0$ the oscillating term $\exp[4it(\alpha^3-\beta
^3)]$ improves the behavior at large $\alpha$ of the integrand.
Consequently, the potential $v(t,p)$ satisfies the condition
\beqs
\label{vconstraint}
\lim_{p_1\to 0}v(t,p)=0\qquad\qquad \mbox{for}\quad t\not=0
\eeqs
in the sense of distribution in $p_2$.

Taking into account the asymptotic behavior in (\ref {lim6}) we see
that differentiation of (\ref {IPJost''}) under the sign of integral
can give a divergent result at $t=0$. The main advantage of the use of
the Fourier transformed Jost solutions $\nu^\pm(t,p|k)$ is that they
give an explicit possibility to remove the rapidly oscillating terms,
given by the `free' part of (\ref {dtnunot0}).  Indeed, let us
introduce the new functions
\beq \label{nutilde}
\widetilde{\nu}^{\,\pm}(t,p|k)=e^{2it[3(k+p_1){\cal L}(p)+2p_1^3]}\,
\nu^{\,\pm}(t,p|k).
\eeq
The kernel of the integral equation for $\widetilde{\nu}^{\,\pm}(t,p|k)$
does not contains a term oscillating in time depending on $\beta$.
In fact, by using the identity
\beq
\label{ident}
3(k+p_1){\cal L}(p)+2p_1^3 +2(\alpha ^3-\beta ^3)=
3(p_1+\alpha ){\cal L}(p+\ell(\alpha )-\ell(\beta) )+2(p_1+\alpha
-\beta )^3 +3{\cal L}(p)(k-\alpha ),
\eeq
we can rewrite (\ref {IPJost''}) in the form
\beqs
\label{IPJost'''}
&&\widetilde{\nu}^{\,\pm} (t,p|k)=\delta ^2(p)-\\
&& \!\int\!\!{d\alpha\,e^{6it{\cal L}(p)(k-\alpha )}
\over \alpha -k\mp i0} \sum_{\sigma =+,-}\!\int\!\!d\beta\, {\sgn
(\alpha -\beta )\over 2} \, \overline{r^{-\sigma }(\alpha ,\beta)}\,
\widetilde{\nu}^{\,\sigma } (t,p+\ell(\alpha )-\ell(\beta )|\beta ). \nonumber
\eeqs
In order to get the evolution equation for $\widetilde{\nu}^{\,\pm} (t,p|k)$
we have to derive this equation with respect to the time and exploit
the  exponential time dependence in the {r.h.s.} by exchanging the
time derivative with the integration over $\alpha$. This can be done
for $t\not=0$ but at $t=0$ it is forbidden because of the following
behavior at large $\alpha$
\beqs
\label{lim7}
&&\lim_{\alpha
\to\infty}\alpha \!\int\!\!d\beta \,{\sgn (\alpha -\beta )\over 2} \,
\overline{r^{-\sigma }(\alpha ,\beta )}\,\widetilde{\nu}^{\,\sigma }(t,
p+\ell(\alpha )-\ell(\beta )|\beta )=
\\ &&-{\delta (p_1)\over 4}\,\sgn p_2 \,v(0,p),\nonumber
\eeqs
that can be easily proved in analogy with (\ref {lim6}) taking into account
that the property $\widetilde{\nu}^\pm(t,p,k)\to\delta^2(p)$ for
$k\to\infty$ holds for any $t$. Note that this limit is time
independent and determined by the initial data $u(0,x,y)$.

In conclusion we rewrite (\ref{IPJost''}) as
\beqs
\label{IPJost3}
& &\widetilde{\nu}^{\,\pm} (t,p|k)=\delta ^2(p)- \\ &
&\!\int\!\!{d\alpha\,e^{6it{\cal L}(p)(k-\alpha )} \over 2(\alpha -k\mp i0)}
\Bigl\{\sum_{\sigma =+,-}\!\int\!\!d\beta \,\sgn (\alpha
-\beta)\,\overline{r^{-\sigma }(\alpha ,\beta )}\, \widetilde{\nu} ^{\,\sigma}
(t,p+\ell(\alpha )-\ell(\beta )|\beta )+\nonumber\\
&&\qquad\qquad\qquad\qquad\qquad\qquad\qquad\qquad\qquad\qquad\qquad
{1\over \alpha\mp i0}\,\delta (p_1)\, \sgn p_2\,v(0,p) \Bigr\}\mp\nonumber\\
&&\qquad\qquad i\pi\,\delta (p_1)\,v(0,p)\,\sgn p_2\,\theta (\mp t{\cal L}(p))
\,{e^{6itk{\cal L}(p)}-1\over k}. \nonumber
\eeqs
The last term compensates the subtracted term since
\beq
\label{izero}
\!\int\!\!{d\alpha\,e^{6it{\cal L}(p)(k-\alpha )} \over (\alpha\mp i0)\,
(\alpha -k\mp i0)} =\mp 2i\pi\theta (\mp t{\cal L}(p))\,{e^{6itk{\cal
L}(p)}-1\over k} .
\eeq

\subsubsection{The first operator in the Lax pair}
The non--stationary Schr\"odinger equation (\ref {Schrodin}) in the Fourier
transformed space (see (\ref {phimu}) and (\ref{munuJ})) reads
\beq
\label{fourierSchrodin}
[{\cal L}(p)-2p_1k]\,\nu (t,p|k)=\!\int\!\!d^2p'\,v(t,p-p')\,\nu (t,p'|k).
\eeq
We want to prove, now, that the function $\nu^\pm $ solution of (\ref
{IPJost2}) obeys (\ref{fourierSchrodin}) with the potential $v(t,p)$ defined
in (\ref {poten:tJ}). Let us rewrite (\ref {IPJost3}) in terms of the
function
\beq
\label{rhoJost'}
\widetilde{\rho}^{\,\pm} (t,p|k)=[{\cal L}(p)-2p_1k]\,
\widetilde{\nu}^{\,\pm} (t,p|k),
\eeq
that, due to (\ref{nutilde}), is related to the function $\rho$ introduced in
(\ref {rhoJost}) by
\beq
\label{rhotilde}
\widetilde{\rho}^{\,\pm}(t,p|k)=e^{2it[3(k+p_1){\cal L}(p)+2p_1^3]}\,
\rho^{\,\pm}(t,p|k).
\eeq
By using the identity
\beqs
{\cal L}(p)-2p_1k={\cal L}(p+\ell(\alpha)
-\ell(\beta))-2(p_1+\alpha-\beta)\beta-2p_1(k-\alpha)
\eeqs
we get
\beqs
& &\widetilde{\rho }^{\,\pm}(t,p|k)=\nonumber\\
& &-\!\int\!\!{d\alpha\,e^{6it{\cal
L}(p)(k-\alpha )} \over 2(\alpha -k\mp i0)} \Bigl\{\sum_{\sigma =+,-}
\!\int\!\!d\beta \,\sgn (\alpha -\beta)\,\overline{r^{-\sigma }(\alpha
,\beta)} \, \widetilde{\rho}^{\,\sigma } (t,p+\ell(\alpha )-\ell(\beta )|\beta
)+\nonumber\\
&&\qquad\qquad\qquad\qquad\qquad\qquad\qquad\qquad\qquad\qquad{1\over
\alpha\mp i0 }\, \delta (p_1)\, |p_2|\,v(0,p) \Bigr\}+\nonumber\\
&&i\pi\,\delta (p_1)\,v(0,p)\,|p_2|\,\theta (\mp t{\cal L}(p))\,
{e^{6itk{\cal L}(p)}-1\over k} -\nonumber\\
&& p_1\!\int\!\!d\alpha\,e^{6it{\cal L}(p)(k-\alpha )}
\Bigl\{\sum_{\sigma =+,-} \!\int\!\!d\beta \,\sgn (\alpha
-\beta)\,\overline{r^{-\sigma }(\alpha ,\beta)}\, \widetilde{\nu} ^{\sigma }
(t,p+\ell(\alpha )-\ell(\beta )|\beta )+\nonumber\\
&&\qquad\qquad\qquad\qquad\qquad\qquad\qquad\qquad\qquad{1\over \alpha\mp i0
}\, \delta (p_1)\,\sgn p_2\,v(0,p) \Bigr\}.
\eeqs
Using again (\ref{rhoJost'}) and (\ref{rhotilde}) and the definition of the
potential in (\ref {poten:tJ}) we have
\beqs
\label{IPJost3'}
& &\rho^{\,\pm} (t,p|k)=v(t,p)-\nonumber\\ &&\!\int\!\!{d\alpha
\over 2(\alpha -k\mp i0 )} \sum_{\sigma =+,-} \!\int\!\!d\beta \,
e^{4it(\alpha^3-\beta ^3 )}\,\sgn (\alpha -\beta)\,\overline{r^{-\sigma
}(\alpha ,\beta )}\, \rho^{\sigma } (t,p+\ell(\alpha )-\ell(\beta )|\beta
).\nonumber
\eeqs
Comparing this integral equation for $\rho^{\pm} (t,p|k)$ with the integral
equation (\ref{IPJost''}) we deduce:
\beq
\label{Schordin''}
\rho^\pm(t,p|k)=\!\int\!\!d^2p'\,v(t,p-p')\,\nu^\pm(t,p'|k)
\eeq
that, due to (\ref {rhoJost}), is just the Fourier transformed non--stationary
Schr\"odinger equation (\ref{fourierSchrodin}).

\subsubsection{The second operator in the Lax pair}
As we already noted, for computing the time derivative at
$t=0$ of $\widetilde{\nu}^{\,\sigma} (t,p|k)$ it is necessary to use the
subtracted equation (\ref{IPJost3}).
We get different left and right
limits
\beqs
\label{IPJost40}
&&{\partial\widetilde{\nu}^{\sigma }\over
\partial t} (\pm 0,p|k)+ \nonumber\\ &&\sum_{\sigma'
=+,-}\!\int\!\!{d\alpha \over 2(\alpha -k-i0\sigma ) } \!\int\!\!d\beta
\,\sgn (\alpha -\beta)\, \overline{r^{-\sigma' }(\alpha ,\beta )}\,
{\partial\widetilde{\nu}^{\sigma '} \over \partial t}(\pm
0,p+\ell(\alpha )-\ell(\beta )|\beta )=\nonumber\\ &&3i{\cal
L}(p)\Bigl\{\int\!\!d\alpha \Bigl[\sum_{\sigma' =+,-}\!\int\!\!d\beta
\,\sgn (\alpha -\beta))\, \overline{r^{-\sigma' }(\alpha ,\beta )}\,\nu
^{\sigma '} (\pm0,p+\ell(\alpha )-\ell(\beta )|\beta )
+\nonumber\\
&&\qquad\qquad\qquad\qquad{1\over \alpha -i0\sigma }\,
\delta (p_1)\,\sgn p_2\,v(0,p) \Bigr]-2i\pi\sigma \,\delta (p_1)\,
\theta (\mp p_2\sigma )\,v(0,p)\Bigr\}.
\eeqs
The formula in curl brackets can be expressed in terms of $v(0,p)$ by
using (\ref {poten5}) and we have
\beqs
\label{IPJost41}
&&{\partial\widetilde{\nu}^{\sigma }\over \partial t} (\pm 0,p|k)+
\nonumber\\ &&\sum_{\sigma' =+,-}\!\int\!\!{d\alpha \over 2(\alpha
-k-i0\sigma ) } \!\int\!\!d\beta \,\sgn (\alpha -\beta)\,
\overline{r^{-\sigma' }(\alpha ,\beta )}\,
{\partial\widetilde{\nu}^{\sigma '} \over \partial t}(\pm
0,p+\ell(\alpha )-\ell(\beta )|\beta )=\nonumber\\
&&-3i{\cal L}(p)\,{v(0,p)\over p_1\pm i0}.
\eeqs
Comparing this integral equation with (\ref{IPJost2}) we deduce that
\beqs
\label{dtnu0}
{\partial\over \partial t} \nu^{\sigma}(\pm0,p|k)&=
&-2i[3(p_1+k){\cal L}(p)+2p_1^3]\, \nu^{\sigma}(\pm0,p|k)- \\
&&3i\!\int\!\!\!\int\!\!d^2p'\,{\cal L}(p')\,{v(0,p')\over p'_1\pm i0}
\, \nu^{\sigma}(\pm0,p-p'|k).\nonumber
\eeqs
For $t\not=0$ the time derivative of $\nu^{\sigma}(t,p|k)$
can be directly computed by using (\ref{IPJost''}). We get (\ref {dtnunot0}),
where the distribution $1/p'_1$ does not need to be specified
as for these $t$ the potential satisfies the condition (\ref{vconstraint}).
Thus we proved that the reconstructed $\nu^{\sigma}(t,p|k)$
satisfies the evolution equation
\beqs
\label{dtnu}
{\partial\over \partial t} \nu^{\sigma}(t,p|k)&= &-2i[3(p_1+k){\cal
L}(p)+2p_1^3]\, \nu^{\sigma}(t,p|k)- \\ &
&3i\!\int\!\!\!\int\!\!d^2p'\,{\cal L}(p')\,{v(t,p')\over p'_1+i0t} \,
\nu^{\sigma}(t,p-p'|k).\nonumber \eeqs

\subsubsection{The evolution version of KPI}
To derive the evolution equation for the reconstructed $v(t,p)$ we
substitute in (\ref{dtnu}) the expansion (\ref {nuJost2}) for
 large $k$ and omit terms of order $1/k^2$. Again it is necessary to
consider separately the case $t=0$. Noting in (\ref {nuJost2})
$\sgn\Im k =\sigma$ and multiplying the result by $k$ we get
\beqs
\label{kp1J''}
&& {1\over p_1+i\sigma  0p_2} {\partial v\over \partial t} (\pm 0,p)=\\
&&6ik{\cal L}(p)\,v(0,p)\left\{{1\over p_1\pm i0}-{1\over p_1+i\sigma
0p_2}\right\}-\nonumber\\
&&i {3{\cal L}(p)^2+6p_1^2{\cal L}(p)+4p_1^4\over (p_1+i\sigma  0p_2)^2}
\,v(0,p)+ \nonumber\\
& &{3i{\cal L}(p)\over p_1+i\sigma
0p_2}\!\int\!\!d^2p'\, {v(0,p-p')\,v(0,p')\over p'_1\pm i0}-3i
\!\int\!\!d^2p'\,{\cal L}(p')\, {v(0,p-p')\,v(0,p')\over
(p'_1\pm i0)\,(p_1-p'_1+i\sigma  0p_2)}.\nonumber
\eeqs
To compensate the terms of first order in $k$ for
$k\to\infty$ we have to chose
\beq
\label{condition1}
\sigma  =\pm\sgn p_2 .
\eeq
It means that the limits $k\to\infty$ and $t\to 0$ commute only if
this condition is satisfied. Notice that in this case the last term in
(\ref {IPJost3}) is absent.
Thus
\beqs
\label{kp1'}
&& {\partial v\over \partial t} (\pm 0,p)=-i{3p_2^2+p_1^4\over p_1\pm
i0}\,v(0,p)+ \\ & &3i\!\int\!\!d^2p'\, [{\cal L}(p)-{\cal L}(p')-{\cal
L}(p-p')]\, {v(0,p-p')\,v(0,p')\over p'_1\pm i0}\nonumber
\eeqs
which due to the identity
\beq
\label{ident2} {\cal L}(p)-{\cal L}(p')-{\cal
L}(p-p')=2p'_1(p'_1-p_1)
\eeq
takes the form
\beq
\label{kp1''}
{\partial v\over \partial t} (\pm 0,p)=-i{3p_2^2+p_1^4\over p_1\pm
i0}\,v(0,p)- 6i\!\int\!\!d^2p'\, (p_1-p'_1)\,
v(0,p-p')\,v(0,p').\nonumber
\eeq
For $t\not=0$ it is not necessary to
use the subtracted form for the reconstructed $v(t,p)$. Finally we
deduce that $v(t,p)$ satisfies the evolution equation
\beq
{\partial v\over \partial t} (t,p)=-i{3p_2^2+p_1^4\over p_1+t
i0}\,v(t,p)- 6i\!\int\!\!d^2p'\, (p_1-p'_1)\,
v(t,p-p')\,v(t,p')\nonumber
\eeq
which is the Fourier transform of the
evolution equation
\beq
\label{kp1'''}
u_t(t,x,y)-6u(t,x,y)u_x(t,x,y)+u_{xxx}(t,x,y)=3\!\int^{x}_{-t\infty}
\!\!dx'\,u_{yy}(t,x',y).
\eeq

\paragraph{Acknowledgements}
A.~P. thanks A.~Fokas and M.~Ablowitz for fruitful discussions
and his colleagues at Dipartimento di Fisica dell'Universit\`a di Lecce for
hospitality.

\end{document}